\documentclass[aps,prc,groupedaddress,showkeys,showpacs,manuscript]{revtex4}
\usepackage{amssymb}
\usepackage{array}
\usepackage{graphicx}
\usepackage{color}
\usepackage{CJKutf8}
\newcommand{\upcite}[1]{\textsuperscript{\textsuperscript{\cite{#1}}}}

\begin{document}

\title{Kinetic Freeze-Out Properties from Transverse Momentum Spectra of Kaon, Pion, and (anti-)Proton Production in U+U Collisions at $\sqrt{s_{NN}}$ = 193 GeV}

\author {Ying Yuan,$\, ^{1}$\footnote{E-mail address: yuany@gxtcmu.edu.cn}}
\address{
1) Mathematics and Physics Teaching and Research Section, College of Pharmacy, Guangxi University of Chinese Medicine, Nanning 530200, China\\}


\begin{abstract}
In the framework of the multi-source thermal model employing the Tsallis distribution, the transverse momentum distributions of kaon, pion, and (anti-)proton production in U+U collisions at $\sqrt{s_{NN}}$ = 193 GeV with varying centrality are investigated. The transverse momentum spectra are appropriately characterized. The dependencies of parameters (average transverse momenta, effective temperature, and entropy index) on event centrality are determined. It is observed that the $q$ parameters increase as the average number of particles participating in the collisions rises, which implies that the nuclear stopping degree elevates with the increase of collision centrality. The $T$ value remains fundamentally consistent for the same particle under different collision parameters, suggesting that the kinetic freezing temperature of particle ejection in this collision system is independent of the collision parameters. However, the q value exceeded the previously determined research range, which might be related to the deformation of the U-nucleus.
\end{abstract}



\maketitle

{\section{Introduction}}

Ultra-relativistic heavy-ion collisions (URHICs) provide an unparalleled experimental avenue to explore strongly interacting matter under extreme temperatures and densities—conditions analogous to those prevailing microseconds after the Big Bang \upcite{Alt1,Sun2,Wang3,Li4,Liu5,Waqas6,Abdulameer7}. A central objective of such studies is to unravel the quantum chromodynamics (QCD) phase structure, particularly the transition from the deconfined quark-gluon plasma (QGP) to the confined hadron gas (HG) \upcite{Arsen8,Li9}. The production mechanisms of hadrons and nuclei in these collisions encode critical signatures of this phase transition, making their investigation pivotal to advancing our understanding of QCD matter. The Relativistic Heavy Ion Collider (RHIC) is uniquely positioned for this research, as it is designed to operate near the critical energy threshold for the hadron-to-QGP phase transition, enabling precise probing of the boundary between these matter states \upcite{Lao10}. 

Over decades, theoretical frameworks such as the thermal model and coalescence model have been developed to interpret hadron production, offering complementary insights into the evolution of collision systems \upcite{Mrowc11,Mrowc12,Bazak13,Liu14,Liu15}. In particular, the study of transport phenomena is of significant importance for comprehending numerous fundamental properties \upcite{Li16}. Among the most informative observables in URHICs are the transverse momentum spectra of produced particles, which serve as a window into the kinetic freeze-out stage—the point at which hadrons cease strong interactions and their final momenta are fixed \upcite{Chen17}. This stage is characterized by key parameters, including the kinetic freeze-out temperature ($T_{kin}$) and average collective flow velocity, which together reflect the thermal excitation and expansion dynamics of the system \upcite{Waqas18,Kumar19}. Notably, the "effective temperature" often extracted directly from $p_{T}$ spectra is not a true thermodynamic temperature but a composite measure encompassing both the system’s excitation degree and the contribution of transverse flow \upcite{Waqas20}. Disentangling these effects to obtain $T_{kin}$—a fundamental marker of the system’s state at freeze-out—remains a core challenge in the field \upcite{Waqas18}.

While extensive studies have probed kinetic freeze-out properties in symmetric, near-spherical heavy-ion systems like Au+Au and Pb+Pb \upcite{Kumar19}, collisions involving highly deformed nuclei such as uranium ($U$) introduce unique complexities that remain underexplored. The strong deformation of U nuclei modulates the geometric overlap of colliding systems across different centralities, potentially altering nuclear stopping, collective flow, and ultimately freeze-out dynamics \upcite{Abdallah21}. 

This study is motivated by the need to leverage this new experimental data to extract reliable kinetic freeze-out temperatures for U+U collisions. We employ the Tsallis distribution—renowned for its ability to capture non-equilibrium features of high-energy collision systems \upcite{Tsallis22,Biro23,Cleymans24} —within the multi-source thermal model, a framework well-suited to describing the multi-component emission of hadrons. By simulating the $p_{T}$ distributions of kaons, pions, and (anti-)protons and comparing our results with STAR’s experimental data \upcite{Abdallah21}, we aim to quantify $T_{kin}$ and its dependence on collision centrality. Beyond extracting this key parameter, our work seeks to lay the groundwork for comparing freeze-out properties between deformed U+U and spherical Au+Au systems, offering insights into how nuclear deformation influences the late-stage evolution of QCD matter.

\vspace{1\baselineskip}

{\section{The model and method}}

The model employed in the current study is the multi-source thermal model \upcite{Liu25,Liu26,Liu27}. In this model, numerous emission sources are formed during high - energy nucleus - nucleus collisions. Various distributions can be utilized to characterize the emission sources and particle spectra, including the Tsallis distribution, the standard (Boltzmann, Fermi - Dirac, and Bose - Einstein) distributions, the Tsallis + standard distributions \upcite{Buyukkilic28,Chen29,Conroy30,Pennini31,Teweldeberhan32,JConroy33}, the Erlang distribution \upcite{Liu25}, etc.The Tsallis distribution can be depicted by two or three standard distributions.

The experimental data of the transverse momentum spectrum of the particles are fitted using the Tsallis distribution, which can account for the temperature fluctuation in several sources to yield an average value. The Tsallis distribution exhibits multiple functional forms \upcite{Tsallis22,Biro23,Cleymans24,Buyukkilic28,Chen29,Conroy30,Pennini31,Teweldeberhan32,JConroy33,Zheng34,Zheng35}, among which the normalized standard momentum distribution relying on the Boltzmann distribution can be expressed as
\begin{equation}
f(p)=\frac{1}{N}\frac{dN}{dp}=Cp^{2}\left\{\left[1\pm\frac{q-1}{T}\left(\sqrt{p^{2}+m_{0}^{2}}-\mu\right)\right]^{\pm\frac{1}{q-1}}\right\}^{-1}.
\label{eq:1}
\end{equation} 
Here, $N$ represents the particle number, $C$ is the normalization constant, $m_{0}$ is the rest mass of the studied particle, $T$ is the temperature that averagely describes several sources (local equilibrium states), $q$ is the entropy index which describes the degree of non-equilibrium among different states, $\mu$ is the chemical potential related to $\sqrt{s_{NN}}$ \upcite{Andronic36}. 

In the rest frame of a considered source, a simplified form of the joint probability is selected: density function of transverse momentum ($p_{\rm T}$) and rapidity ($y$) \upcite{Lao10},
\begin{equation}
f(p_{\rm T},y)\propto{\frac{d^{2}N}{dydp_{T}}}={\frac{gV}{{(2\pi)}^{2}}}{p_{T}}{\sqrt{p_{\rm T}^{2}+m_{\rm 0}^{2}}}{\cosh{y}}\times{{[1\pm{\frac{q-1}{T}}({\sqrt{p_{\rm T}^{2}+m_{\rm 0}^{2}}}{\cosh{y}}-{\mu})]}^{\pm{\frac{q}{q-1}}}}.
\label{eq:2}
\end{equation}
Here, $g$ is the degeneracy factor, $V$ is the volume of emission sources. In the RHIC energy region, $\mu$ is very small, the $\pm$ in the formula takes the plus sign. The values of $T$, $q$ and $V$ are obtained from reproducing the particle spectra, where $T$ and $q$ are independently fitted for the studied particle, and $V$ is related to other parameters.

The Monte Carlo distribution generating method is adopted to obtain $p_{\rm T}$. Let $r_{\rm 1}$ denote the random numbers uniformly distributed in $[0,1]$. A series of values of $p_{\rm T}$ can be acquired through:
\begin{equation}
\int_{0}^{p_{T}}f_{p_{T}}(p_{T})dp_{T}< r_{1}< \int_{0}^{p_{T}+dp_{T}}f_{p_{T}}(p_{T})dp_{T}.
\label{eq:3}
\end{equation}
Here, $f_{p_{T}}$ is the transverse momentum probability density function, which is an alternative representation of the Tsallis distribution as follows:
\begin{equation}
f_{P_{T}}\left ( p_{T}\right )=\frac{1}{N}\frac{dN}{dp_{T}}=\int_{y_{min}}^{y_{max}}f\left ( p_{T},y\right )dy.
\label{eq:4}
\end{equation}
where $y_{\text{max}}$ and $y_{\text{min}}$ are the maximum and minimum rapidity, respectively.

Under the assumption of isotropic emission in the source rest frame, the Monte Carlo method is used to obtain the polar angle:
\begin{equation}
\theta=2{\arcsin\sqrt{r_{2}}}.
\label{eq:5}
\end{equation}
Thus, a series of values of momentum and energy can be obtained based on the momentum $p=\frac{p_{T}}{\sin\theta}$ and the energy $E={\sqrt{p^{2}+m_{\rm 0}^{2}}}$. Therefore, the corresponding values of rapidity can be derived according to the definition of rapidity. 

\vspace{1\baselineskip}

{\section{Results and discussion}}

{\subsection{Transverse momentum spectra}}

\begin{figure}
\setlength{\abovedisplayskip}{-0.5cm}
\includegraphics[angle=0,width=16.6cm]{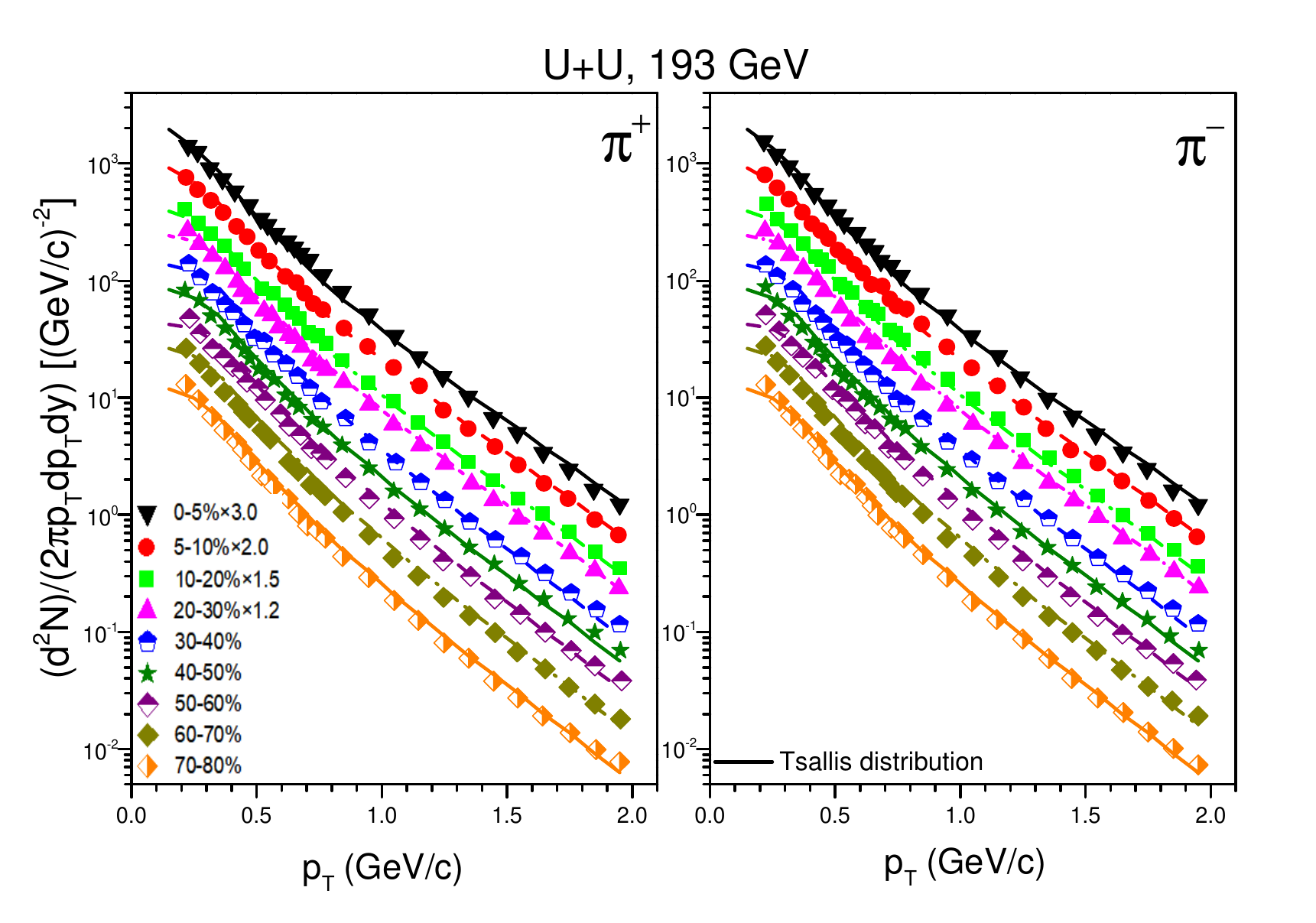}
\caption{Transverse momentum spectra of $\pi^{+}$ and $\pi^{-}$ are calculated at mid-rapidity ($\left | y \right | < 0.1$) in U+U collisions at $\sqrt{s_{NN}}$=193 GeV for $0-5\%$, $5-10\%$, $10-20\%$, $20-30\%$, $30-40\%$, $40-50\%$, $50-60\%$, $60-70\%$ and $70-80\%$ centrality. The theoretical calculation results based on the Tsallis distribution are represented by lines. Experimental data taken from the STAR Collaboration \upcite{Abdallah21} are represented by the symbols.} \label{fig1}
\end{figure}

Fig.\ \ref{fig1} depicts the transverse momentum spectra within nine centrality classes in U+U collisions at $\sqrt{s_{NN}}$=193 GeV at mid-rapidity ($\vert{y}\vert$$<$0.1) for $\pi^{+}$ and $\pi^{-}$. There exist nine centrality classes, representing ranges of $0-5\%$, $5-10\%$, $10-20\%$, $20-30\%$, $30-40\%$, $40-50\%$, $50-60\%$, $60-70\%$ and $70-80\%$ respectively. The symbols denote the experimental data from the STAR Collaboration \upcite{Abdallah21}. The lines represent our calculated results fitted by utilizing the Tsallis distribution based on equation (2) in the mid-rapidity region. The values of the relevant parameters $T$ and $q$ are presented in Table I, along with $\chi^{2}/dof$ (where $\chi^{2}$ is the chi-square value and $dof$ is the number of degrees of freedom). It is observed that the calculations from the Tsallis distribution are in good agreement with the experimental data.

\begin{figure}
\setlength{\abovedisplayskip}{-0.5cm}
\includegraphics[angle=0,width=16.6cm]{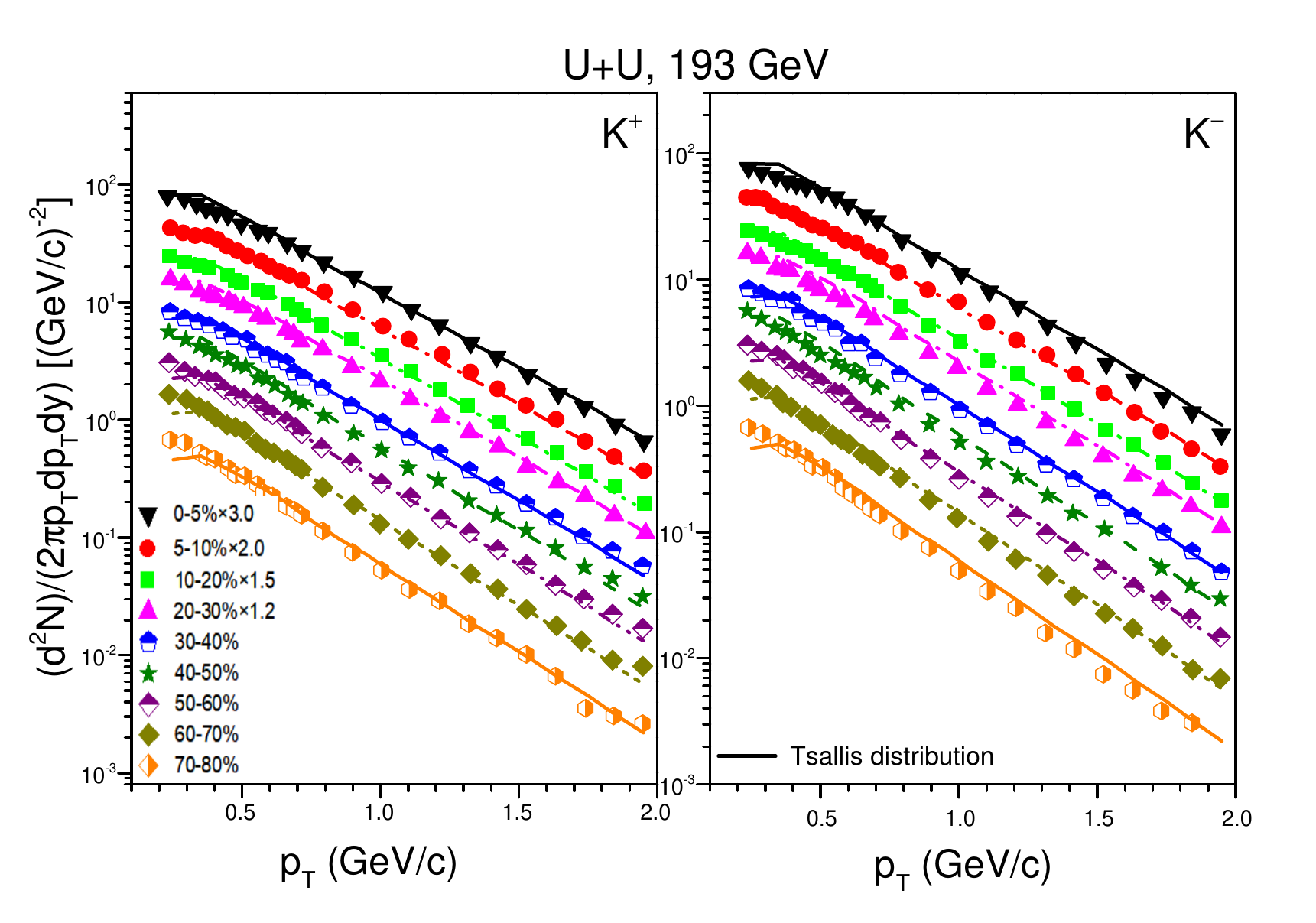}
\caption{Transverse momentum spectra of $K^{+}$ and $K^{-}$ are calculated at mid-rapidity ($\left | y \right | < 0.1$) in U+U collisions at $\sqrt{s_{NN}}$=193 GeV for $0-5\%$, $5-10\%$, $10-20\%$, $20-30\%$, $30-40\%$, $40-50\%$, $50-60\%$, $60-70\%$ and $70-80\%$ centrality. The theoretical calculation results based on the Tsallis distribution are represented by lines. Experimental data taken from the STAR Collaboration \upcite{Abdallah21} are represented by the symbols.} \label{fig2}
\end{figure}

Fig.\ \ref{fig2} illustrates the transverse momentum spectra across nine centrality classes in U+U collisions at $\sqrt{s_{NN}}$=193 GeV at mid-rapidity ($\vert{y}\vert$$<$0.1) for $K^{+}$ and $K^{-}$. The lines are the results computed from the Tsallis distribution. The symbols represent the experimental data of the STAR Collaboration \upcite{Abdallah21}. The values of the relevant parameters $T$ and $q$ are given in Table II, along with $\chi^{2}/dof$. It is found that the calculations of the Tsallis distribution are in good accordance with the experimental data.

\begin{figure}
\setlength{\abovedisplayskip}{-0.5cm}
\includegraphics[angle=0,width=16.6cm]{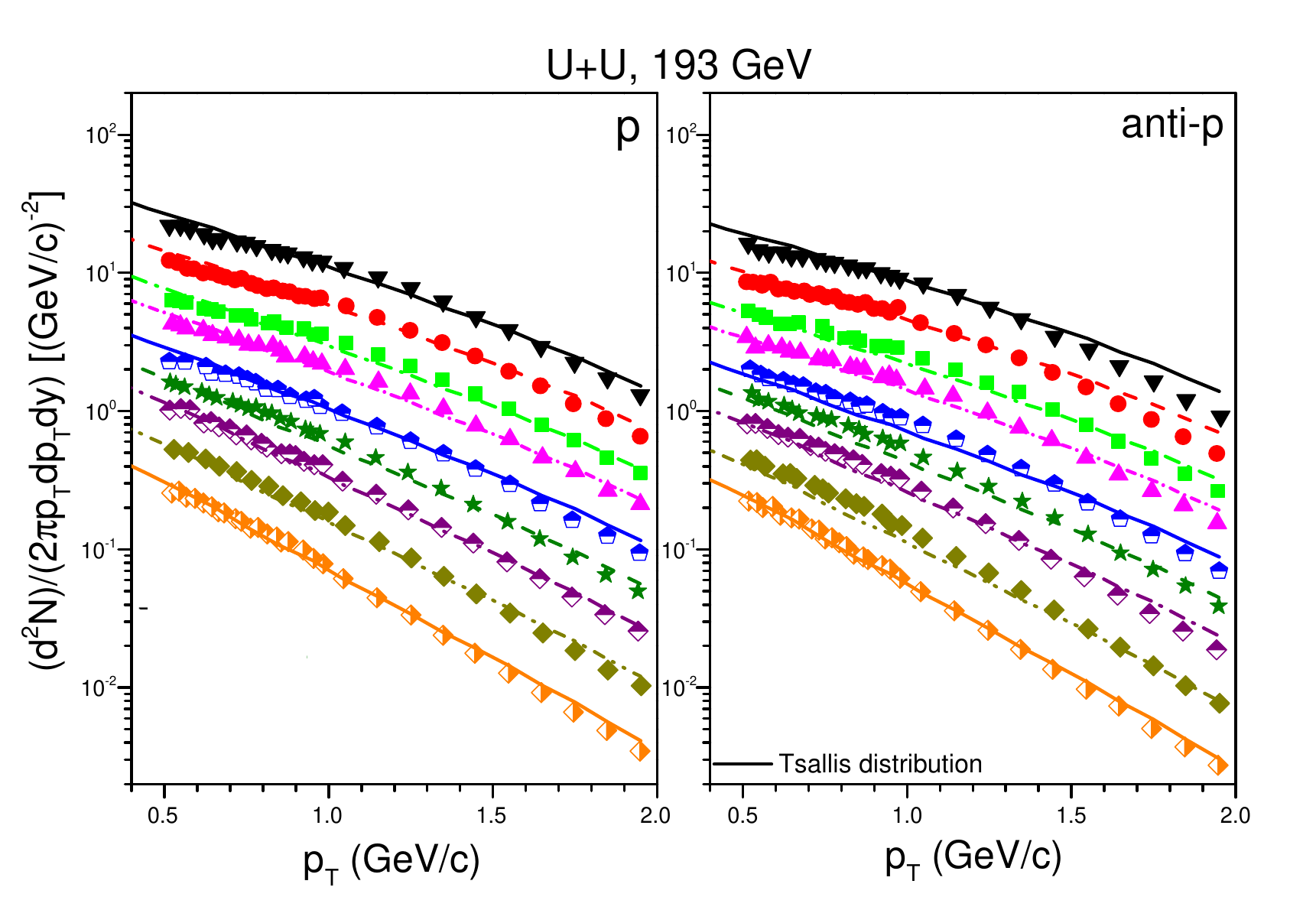}
\caption{The transverse momentum spectra of $p$ and $\bar{p}$ are computed at mid-rapidity ($\left | y \right | < 0.1$) in $U+U$ collisions at $\sqrt{s_{NN}}$=193 GeV for centrality intervals of $0-5\%$, $5-10\%$, $10-20\%$, $20-30\%$, $30-40\%$, $40-50\%$, $50-60\%$, $60-70\%$ and $70-80\%$. The theoretical calculation results based on the Tsallis distribution are represented by lines. The experimental data sourced from the STAR Collaboration \upcite{Abdallah21} are denoted by the symbols.} \label{fig3}
\end{figure}

Fig.\ \ref{fig3} presents the transverse momentum spectra for nine centrality classes in U+U collisions at $\sqrt{s_{NN}}$=193 GeV, measured at mid-rapidity ($\vert{y}\vert$$<$0.1) for protons ($p$) and antiprotons ($\bar{p}$). The lines are the results obtained from the Tsallis distribution. The symbols represent the experimental data of the STAR Collaboration \upcite{Abdallah21}. The values of the related parameters $T$ and $q$ are provided in Table III, along with $\chi^{2}/dof$.  It is noted that the calculations of the Tsallis distribution are in good conformity with the experimental data.

\begin{table}
	\caption{Values of $T$, $q$, and $\chi^{2}/dof$ corresponding to the curves in U+U collisions for $\pi^{+}/\pi^{-}$ at $\sqrt{s_{NN}}$=193 GeV for $0-5\%$, $5-10\%$, $10-20\%$, $20-30\%$, $30-40\%$, $40-50\%$, $50-60\%$, $60-70\%$ and $70-80\%$ centrality.} \label{Table1}
	\begin{tabular}{ p{2cm}<{\centering} p{1.5cm}<{\centering} p{2cm}<{\centering} p{3cm}<{\centering} p{3cm}<{\centering} p{3cm}<{\centering}}
		\hline \hline
		Figure & Type 1 & Type 2 & T (GeV) & q & $\chi^{2}/dof$ \\
		\hline
		Fig. 1 & $\pi^{+}/\pi^{-}$ & 0-5\%   & $0.075\pm0.042$ & $1.288\pm0.100$ & 0.010 \\
		       &       & 5-10\%  & $0.077\pm0.018$ & $1.280\pm0.068$ & 0.004 \\
	    	   &       & 10-20\% & $0.079\pm0.005$ & $1.272\pm0.014$ & 0.006 \\
		       &       & 20-30\% & $0.081\pm0.035$ & $1.268\pm0.160$ & 0.010 \\
		       &       & 30-40\% & $0.080\pm0.008$ & $1.258\pm0.357$ & 0.006 \\
		       &       & 40-50\% & $0.079\pm0.008$ & $1.263\pm0.356$ & 0.008 \\
		       &       & 50-60\% & $0.081\pm0.008$ & $1.258\pm0.339$ & 0.013 \\
		       &       & 60-70\% & $0.079\pm0.037$ & $1.258\pm0.137$ & 0.032 \\
		       &       & 70-80\% & $0.078\pm0.032$ & $1.258\pm0.086$ & 0.007 \\
		\hline \hline
	\end{tabular}
\end{table}

\begin{table}
	\caption{Values of $T$, $q$, and $\chi^{2}/dof$ corresponding to the curves in U+U collisions for $K^{+}/K^{-}$ at $\sqrt{s_{NN}}$=193 GeV for $0-5\%$, $5-10\%$, $10-20\%$, $20-30\%$, $30-40\%$, $40-50\%$, $50-60\%$, $60-70\%$ and $70-80\%$ centrality.} \label{Table2}
	\begin{tabular}{ p{2cm}<{\centering} p{1.5cm}<{\centering} p{2cm}<{\centering} p{3cm}<{\centering} p{3cm}<{\centering} p{3cm}<{\centering}}
		\hline \hline
		Figure & Type 1 & Type 2 & T (GeV) & q & $\chi^{2}/dof$ \\
		\hline
		Fig. 2 & $K^{+}/K^{-}$ & 0-5\%   & $0.098\pm0.030$ & $1.308\pm0.269$ & 0.016 \\
		&       & 5-10\%  & $0.100\pm0.015$ & $1.300\pm0.068$ & 0.005 \\
		&       & 10-20\% & $0.100\pm0.048$ & $1.292\pm0.102$ & 0.006 \\
		&       & 20-30\% & $0.102\pm0.009$ & $1.288\pm0.356$ & 0.021 \\
		&       & 30-40\% & $0.100\pm0.015$ & $1.284\pm0.068$ & 0.006 \\
		&       & 40-50\% & $0.098\pm0.012$ & $1.278\pm0.060$ & 0.026 \\
		&       & 50-60\% & $0.100\pm0.013$ & $1.278\pm0.056$ & 0.021 \\
		&       & 60-70\% & $0.100\pm0.018$ & $1.270\pm0.004$ & 0.038 \\
		&       & 70-80\% & $0.102\pm0.013$ & $1.260\pm0.076$ & 0.023 \\
		\hline \hline
	\end{tabular}
\end{table}

\begin{table}
	\caption{Values of $T$, $q$, and $\chi^{2}/dof$ corresponding to the curves in U+U collisions for $p$ and $\bar{p}$ at $\sqrt{s_{NN}}$=193 GeV for $0-5\%$, $5-10\%$, $10-20\%$, $20-30\%$, $30-40\%$, $40-50\%$, $50-60\%$, $60-70\%$ and $70-80\%$ centrality.} \label{Table3}
	\begin{tabular}{ p{2cm}<{\centering} p{1.5cm}<{\centering} p{2cm}<{\centering} p{3cm}<{\centering} p{3cm}<{\centering} p{3cm}<{\centering}}
		\hline \hline
		Figure & Type 1 & Type 2 & T (GeV) & q & $\chi^{2}/dof$ \\
		\hline
		Fig. 3 & $p$ & 0-5\%   & $0.102\pm0.014$ & $1.418\pm0.485$ & 0.011 \\
		&       & 5-10\%  & $0.102\pm0.023$ & $1.414\pm0.072$ & 0.007 \\
		&       & 10-20\% & $0.100\pm0.042$ & $1.409\pm0.132$ & 0.021 \\
		&       & 20-30\% & $0.102\pm0.021$ & $1.394\pm0.111$ & 0.022 \\
		&       & 30-40\% & $0.102\pm0.010$ & $1.384\pm0.081$ & 0.010 \\
		&       & 40-50\% & $0.100\pm0.016$ & $1.370\pm0.055$ & 0.009 \\
		&       & 50-60\% & $0.100\pm0.003$ & $1.345\pm0.037$ & 0.004 \\
		&       & 60-70\% & $0.100\pm0.020$ & $1.335\pm0.064$ & 0.008 \\
		&       & 70-80\% & $0.100\pm0.018$ & $1.305\pm0.081$ & 0.006 \\
		\hline
		Fig. 3 & $\bar{p}$ & 0-5\%  & $0.103\pm0.022$ & $1.442\pm0.081$ & 0.042 \\
		&       & 5-10\%  & $0.103\pm0.012$ & $1.435\pm0.140$ & 0.022 \\
		&       & 10-20\% & $0.103\pm0.025$ & $1.425\pm0.165$ & 0.015 \\
		&       & 20-30\% & $0.103\pm0.020$ & $1.415\pm0.149$ & 0.015 \\
		&       & 30-40\% & $0.102\pm0.005$ & $1.400\pm0.076$ & 0.021 \\
		&       & 40-50\% & $0.102\pm0.013$ & $1.375\pm0.088$ & 0.021 \\
		&       & 50-60\% & $0.102\pm0.007$ & $1.355\pm0.031$ & 0.019 \\
		&       & 60-70\% & $0.102\pm0.008$ & $1.325\pm0.282$ & 0.025 \\
		&       & 70-80\% & $0.102\pm0.002$ & $1.295\pm0.079$ & 0.027 \\
		\hline \hline
	\end{tabular}
\end{table}

Under normal conditions, the $q$ value value lies between 1.0 and 1.2; however, the $q$ values in the above tables exceed this range \upcite{Cleymans37,Cleymans38}. Given that the U nucleus is the most deformed nucleus, the correction for nuclear deformation is not considered in the current Tsallis distribution, thus resulting in a relatively large $q$ value. The $T$ value remains essentially consistent under different collision centrality, which is attributable to the fact that the orientation of the U nucleus is isotropic in the calculation.

{\subsection{Average transverse momenta distributions}}

Figure 4 shows the variation of $\left \langle p_{T}  \right \rangle$ with $\left \langle N_{part}  \right \rangle$ at mid-rapidity ($\left | y \right | < 0.1$) for $\pi^{+}$, $K^{+}$ and $p$ particles in U+U collisions at $\sqrt{s_{NN}}$=193 GeV. The red solid circles represent the experimental data from the STAR Collaboration \upcite{Abdallah21}, and the black diamonds are the calculations from the Tsallis distribution. The calculations can be derived by
\begin{equation}
\left \langle {p_{T}}\right \rangle=\frac{\sum {p_{T1}}{\alpha }}{\sum {\alpha} }.
\label{eq:6}
\end{equation}
Here, ${p_{T1}}$ is the value of transverse momentum corresponding to the experimental data, and $\alpha$ is the value of 
$\frac{d^{2}N}{{N_{event}}2\pi {p_{T}}d{p_{T}}dy}$ that corresponds to the ${p_{T1}}$.
It is found that the experimental results can be described within the margin of error. The values of $\left \langle p_{T}  \right \rangle$ increase gradually with the increase of number of participating nucleons, and they are listed in Table IV. In other words, the greater the intensity of the collision, the higher the transverse momentum of the emitted particles.

\begin{figure}[htbp] \centering
	\setlength{\abovedisplayskip}{-0.5cm}
	\includegraphics[angle=0,width=16.6cm]{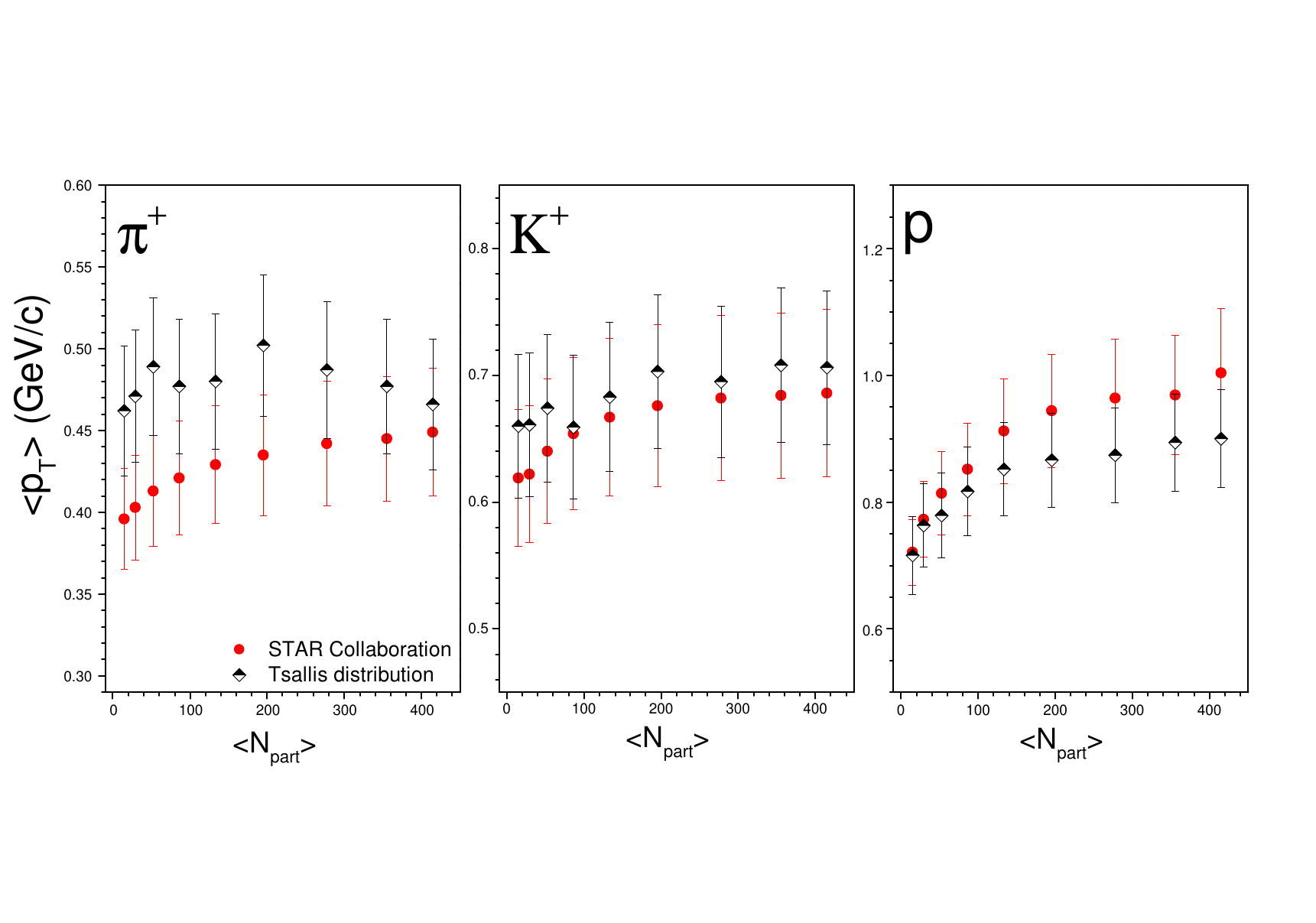}
	\caption{The $\left \langle p_{T}  \right \rangle$ as a function of $\left \langle N_{part}  \right \rangle$ at mid-rapidity ($\left | y \right | < 0.1$) of $\pi^{+}$, $K^{+}$ and $p$ for $U+U$ collisions at $\sqrt{s_{NN}}$=193 GeV. The red solid circles represent data collected by the STAR Collaboration \upcite{Abdallah21}. The black diamonds represent the calculations based on the Tsallis distribution. } \label{fig4}
\end{figure}

\begin{table}
	\caption{Values of $\left \langle p_{T}  \right \rangle$ in GeV/c within mid-rapidity ($\left | y \right | < 0.1$) of $\pi^{+}$, $\pi^{-}$, $K^{+}$, $K^{-}$, $p$ and $\bar{p}$ for U+U collisions at $\sqrt{s_{NN}}$=193 GeV using the Tsallis distribution.} \label{Table4}
	\begin{tabular}{ p{1.8cm}<{\centering} p{2.3cm}<{\centering} p{2.3cm}<{\centering} p{2.3cm}<{\centering} p{2.3cm}<{\centering} p{2.3cm}<{\centering} p{2.3cm}<{\centering}}
		\hline \hline
		Centrality & $\pi^{+}$ & $\pi^{-}$ & $K^{+}$ & $K^{-}$ & $p$ & $\bar{p}$ \\
		\hline
		0-5\%   & $0.466\pm 0.040$ & $0.466\pm 0.040$ & $0.706\pm0.061$ & $0.706\pm0.061$ & $0.900\pm0.077$ & $0.938\pm0.081$ \\
		5-10\%  & $0.477\pm 0.041$ & $0.477\pm 0.041$ & $0.708\pm0.061$ & $0.708\pm0.061$ & $0.894\pm0.077$ & $0.929\pm0.080$ \\
		10-20\% & $0.487\pm 0.042$ & $0.487\pm 0.042$ & $0.695\pm0.060$ & $0.695\pm0.060$ & $0.874\pm0.075$ & $0.916\pm0.079$ \\
		20-30\% & $0.502\pm 0.043$ & $0.502\pm 0.043$ & $0.703\pm0.060$ & $0.703\pm0.060$ & $0.866\pm0.074$ & $0.902\pm0.078$ \\
		30-40\% & $0.480\pm 0.041$ & $0.480\pm 0.041$ & $0.683\pm0.059$ & $0.683\pm0.059$ & $0.852\pm0.073$ & $0.875\pm0.075$ \\
		40-50\% & $0.477\pm 0.041$ & $0.477\pm 0.041$ & $0.659\pm0.057$ & $0.659\pm0.057$ & $0.817\pm0.070$ & $0.838\pm0.072$ \\
		50-60\% & $0.489\pm 0.042$ & $0.489\pm 0.042$ & $0.674\pm0.058$ & $0.674\pm0.058$ & $0.779\pm0.067$ & $0.808\pm0.069$ \\
		60-70\% & $0.471\pm 0.041$ & $0.471\pm 0.041$ & $0.661\pm0.057$ & $0.661\pm0.057$ & $0.763\pm0.066$ & $0.762\pm0.066$ \\
		70-80\% & $0.462\pm 0.040$ & $0.462\pm 0.040$ & $0.660\pm0.057$ & $0.660\pm0.057$ & $0.716\pm0.062$ & $0.714\pm0.061$ \\
		\hline \hline
	\end{tabular}
\end{table}

{\subsection{Dependence of parameters on number of participating nucleons}}

\begin{figure}
\setlength{\abovedisplayskip}{-0.5cm}
\includegraphics[angle=0,width=16.6cm]{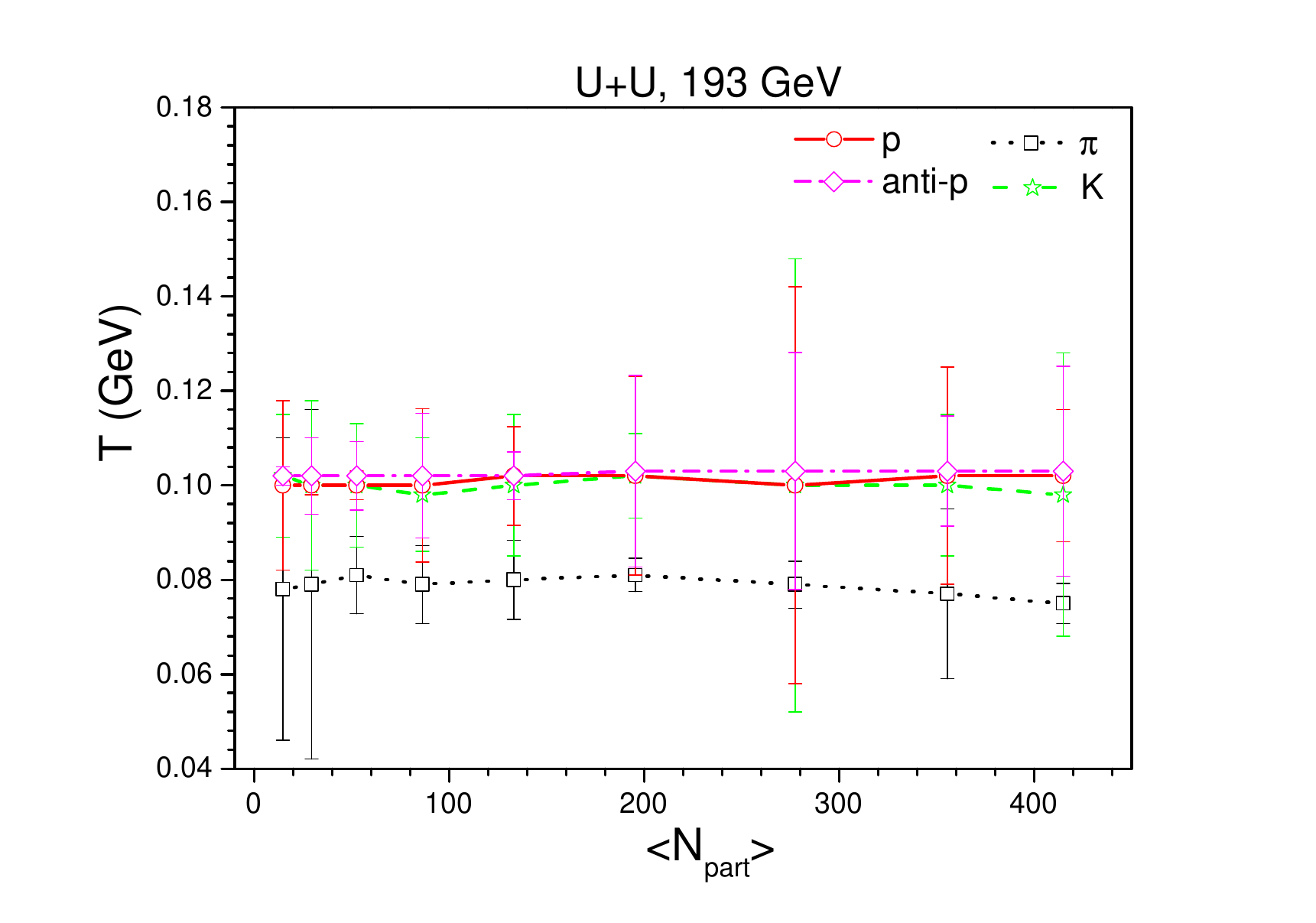}
\caption{Dependence of $T$ on the average number of participants for $\pi^{+}/\pi^{-}$, $K^{+}/K^{-}$, $p$ and $\bar{p}$ in events with different centrality intervals. The symbols represent the parameter values listed in Table I, II and III.} \label{fig5}
\end{figure}

\begin{figure}
\setlength{\abovedisplayskip}{-0.5cm}
\includegraphics[angle=0,width=16.6cm]{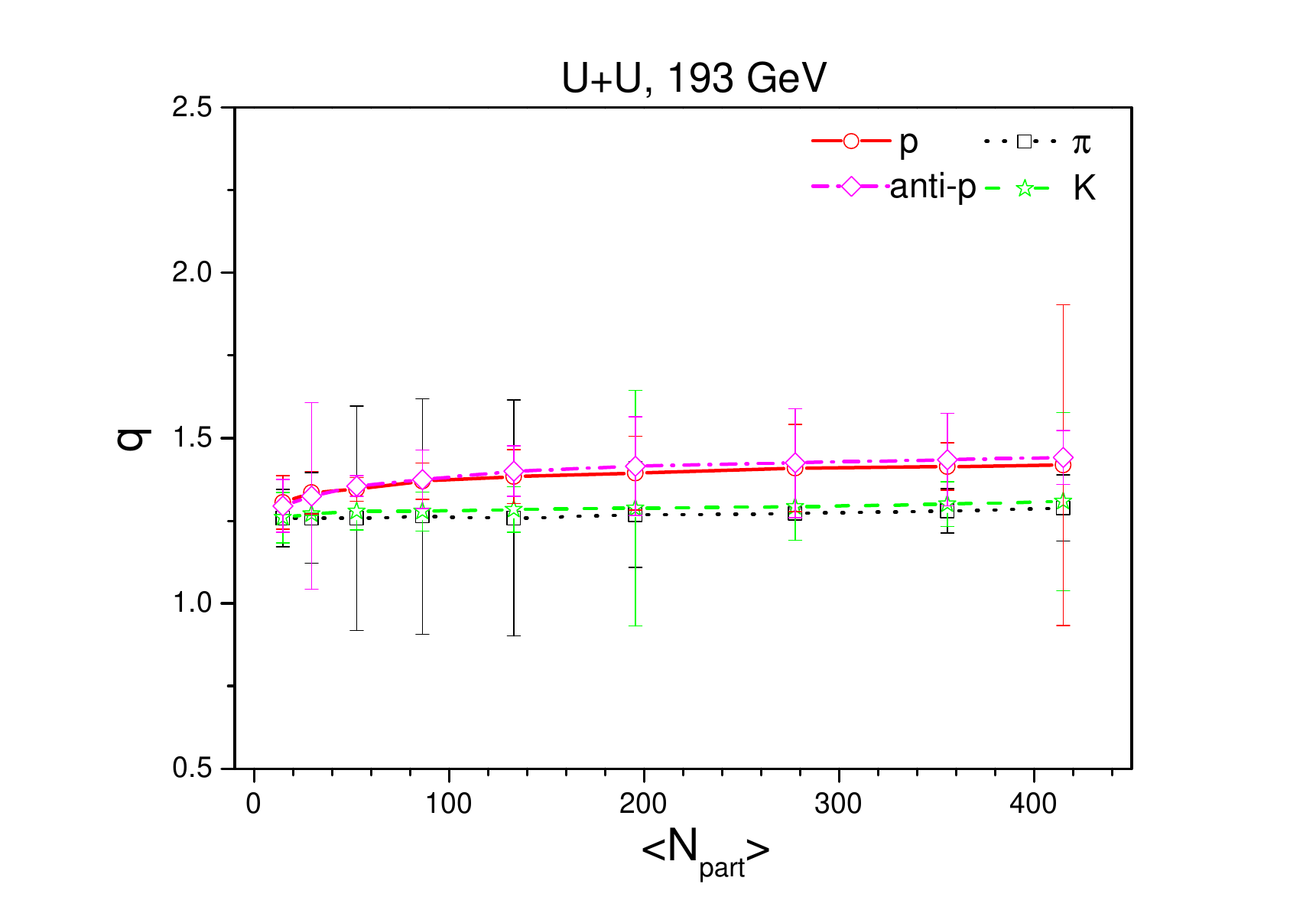}
\caption{Dependence of $q$ on the average number of participants for $\pi^{+}/\pi^{-}$, $K^{+}/K^{-}$, $p$ and $\bar{p}$ in events with different centrality intervals. The symbols represent the parameter values listed in Table I, II and III.} \label{fig6}
\end{figure}

Fig.\ \ref{fig5} and Fig.\ \ref{fig6} illustrate the variation trends of parameters ($T$ and $q$) with the average number of participants for $\pi$ ($\pi^{+}/\pi^{-}$), $K$ ($K^{+}/K^{-}$), $p$ and $\bar{p}$ generated in $U+U$ collisions at $\sqrt{s_{NN}}$=193 GeV in the mid-rapidity region ($\left | y \right | < 0.1$). The symbols denote the parameter values extracted from Figures 1-3 and listed in Tables I-III. 

From Fig.\ \ref{fig5} and Fig.\ \ref{fig6}, it can be observed that the $T$ value remains relatively stable for the same particle, whereas the $q$ value increases as the collision centrality rises. In high-energy experiments, the tip-to-tip collisions and body-to-body collisions of the $U$ nucleus exhibit distinctly different characteristics. However, a slight mass hierarchy phenomenon was also observed: protons (and anti-protons) exhibited the highest $T$ value, followed by $K$ mesons, with $\pi$ mesons displaying the lowest. The $q$ value also shows a dependence on the mass of the particle. This trend can be attributed to the fact that heavier particles, such as protons, more efficiently acquire kinetic energy from the system's collective expansion, leading to a harder transverse momentum spectrum and consequently a higher fitted effective temperature $T$. The closer to the center of the collision, the higher the temperature generated by the fireball, the greater the energy density, and the more significant the non-equilibrium characteristics of the system are usually, thus having a higher entropy value.

\vspace{1\baselineskip}

{\section{Summary and Outlook}}

In conclusion, within the centrality classes of $0-5\%$, $5-10\%$, $10-20\%$, $20-30\%$, $30-40\%$, $40-50\%$, $50-60\%$, $60-70\%$, and $70-80\%$ in $U+U$ collisions at $\sqrt{s_{NN}} = 193$ GeV,the transverse momentum spectra of $\pi^{\pm}$, K$^{\pm}$, and p(${\bar{p}}$) in mid-rapidity region ($|y| < 0.1$) were measured. Additionally, other observable extracted from the transverse momentum spectra, such as the average transverse momentum ($ \langle p_{T} \rangle$), and the relationships regarding effective temperature and entropy are presented as functions of collision centrality. The experimental results from the STAR Collaboration \upcite{Abdallah21} were analyzed using the Tsallis distribution. It was found that the theoretical calculation results can effectively describe the experimental data, and the function of $ \langle p_{T} \rangle$ depends on centrality. The T value remains basically consistent for the same particle under different collision parameters. The q value increases as the collision parameters decrease, yet it exceeds the previously determined research scope. Subsequently, in-depth research will examine Tsallis distribution corrections in deformed nuclei. Further studies on the kinetic freeze-out temperature and collision time evolution are still needed.

\vspace{1\baselineskip}

{\section*{Data Availability}}

The data used to support the findings of this study are included within the article and are cited at relevant places within the text as references.

{\section*{Conflict of Interests}}

The author declare that there is no conflict of interests regarding the publication of this paper.

{\section*{Acknowledgements}}

 This work was supported by the Fund for Less Developed Regions of the National Natural Science Foundation of China under Grant No.12365017, the Natural Science Foundation of Guangxi Zhuangzu Autonomous Region of China under Grant No. 2021GXNSFAA196052, the Introduction of Doctoral Starting Funds of Scientific Research of Guangxi University of Chinese Medicine under Grant No.2018BS024.

\end{document}